# Fractal plasmonic metamaterials for subwavelength imaging

Xueqin Huang[1], Dexin Ye[2], Shiyi Xiao[1], Jiangtao Huangfu[2], Zhiyu Wang[2], Lixin Ran[2], and Lei Zhou[1*]

[1] *Surface Physics Laboratory (State Key Laboratory) and Physics Department, Fudan University, Shanghai 200433, P. R. China*

[2] *The Electromagnetic Academy at Zhejiang University, Zhejiang University, Hangzhou 310027, P. R. China*

**Abstract**

We show that a metallic plate with fractal-shaped slits can be homogenitized as a plasmonic metamaterial with plasmon frequency dictated by the fractal geometry. Owing to the *all-dimensional subwavelength* nature of the fractal pattern, our system supports both transverse-electric and transverse-magnetic surface plasmons. As a result, this structure can be employed to focus light sources with *all-dimensional subwavelength* resolutions and enhanced field strengths. Microwave experiments reveal that the best achievable resolution is only $\lambda/15$, and simulations demonstrate that similar effects can be realized at infrared frequencies with appropriate designs.

Surface plasmon polaritons (SPPs) are elementary electromagnetic (EM) excitations bounded at metal/dielectric interfaces, and attracted considerable attention recently [1]. For a natural material, its plasmon frequency ($\omega_p$) is fixed by the electron density, so that many SPP-based applications only work at a single frequency. Recently, people showed that Bragg scatterings can modulate the SPPs significantly, and found high optical transmissions in a silver film drilled with periodic holes [2-3]. However, the Bragg mechanism can only fold the SPP bands into the first Brillouin zone, but can not change the $\omega_p$ of a material. In 2004, Pendry *et al.* demonstrated that a metallic plate with periodic square holes can mimic a plasmonic material in terms of SPP properties, with effective $\omega_p$ being the waveguide cut-off frequency of the hole [4-5]. This opens up a way to design artificial plasmonic metamaterials at any desired frequencies. However, to make the idea work, one has to fill the holes with *high-index materials* [4-5] which is not easy to realize in practice, particularly at higher frequencies. Very recently, Shin *et al.* [6] showed that high-index insertion is not necessary if the square holes are replaced by closely packed narrow rectangle holes with cross sections $b_x \times b_y$ ($b_x \ll b_y$). Indeed, such a hole has a cut-off wavelength $\lambda = 2b_y$ much longer than the periodicity along $x$ direction (~$b_x$ since the holes are closely packed) [6]. However, since such holes are subwavelength *only* along *one* direction, we will show that the generated SPPs on such structures *only* have transverse-magnetic (TM) polarization travelling along *one* ($x$) direction. This limitation restricts the applications of such structures in many cases.

Here, we demonstrate that a metallic plate drilled with fractal-shaped slits exhibits SPPs with $\omega_p$ dictated by the fractal geometry. Without using high-index insertions [4-5] and distinct from the narrow rectangle hole case [6], here the fractal pattern is *subwavelength along all dimensions* at resonance. Therefore, we found such a system can be homogenitized as a plasmonic metamaterial to support TM and transverse-electric (TE) polarized SPPs *simultaneously*. We further show by both experiments and finite-difference-time-domain (FDTD) simulations that our structure works as a *super* lens to focus light sources with *all dimensional subwavelength* resolutions (best achievable resolution ~ $\lambda/15$), with physical mechanism different from other lenses discussed previously [7-8].

Figure 1 schematically shows our structure — a $0.5 \mu m$-thick silver film caved with 4-level fractal slits arranged periodically with a lattice constant $a = 1\mu m$. We performed FDTD simulations [9] with dispersive $\varepsilon_{Ag}$ given by [10] to calculate the SPP band structure of the designed system. Since this system shows no *x-y* symmetry, we depicted in Fig. 2(a) and (b) the SPP bands along $\Gamma \to X$ and $\Gamma \to X'$ directions, respectively, with $X = [\pi/a, 0, 0]$, $X' = [0, \pi/a, 0]$. In each panel, we found two SPP bands well below the light lines (solid pink line), which bend drastically at two frequencies ($f_{P1} = 41$THz, $f_{P2} = 78.7$ THz). We employed the attenuated total reflection technique to identify the polarization properties of the SPP bands. With $k_x$ fixed as $\pi/a$, we shine evanescent waves with different polarizations on the structure, and depicted in Fig. 2(c) the transmission spectra for $\vec{E} \parallel \hat{x}$ (red circles) and $\vec{E} \parallel \hat{y}$ (blue line) polarizations. By comparing Fig. 2(a) with 2(c), we found that the lower SPP band in Fig. 2(a) is apparently TM-like, since a TM-polarized evanescent wave can excite this mode but a TE one cannot. Similarly, we identified the polarizations of all SPP bands and then labeled them correspondingly on the figure. This is intriguing at first sight, since a complex structure at resonance usually exhibits complicated local field distribution, making it difficult to identify the eigenmode polarizations. We can understand this point in the spirit of "metamaterial". As the probing wavelength is much longer than the unit-cell size, one can perform field average to homogenitize the complex structure as an effective medium. Thanks to the high symmetry of the fractal geometry, the averaged field exhibits well-defined polarization characteristic so that identifying the SPP mode polarization is possible.

The crucial advantage of our structure is clear. Whereas a flat Ag film supports only TM polarized SPP [11], our system supports *simultaneously* TM and TE polarized SPPs related to each resonance. We also studied a metal plate with narrow rectangular holes [6], and depicted its SPP dispersions in the insets to Fig. 2(a)-(b). Indeed, this structure exhibits a SPP band even without high-index insertions. However, in sharp contrast to our case, such a system supports only a single TM-like SPP band traveling along *x* direction [Fig. 2(a)]. This is because the rectangle shape is subwavelength along only one (*x*) direction, and therefore, the SPP band along *y* direction can *not* be formed since the subwavelength condition is not satisfied [4-5]. In contrast, our fractal pattern is *subwavelength* along *all directions* and possesses *multiple* resonances, so that for each resonance, SPP bands along both *x* and *y* directions can be formed (see Fig. 2). FDTD simulations show that the plasmon frequencies can be changed via adjusting the fractal geometry and scaling the unit cell size [12]. Therefore, we can in principle design a plasmoinc metamaterial at any desired frequency.

One can employ our structures to realize many SPP-based applications. It was shown both theoretically [13] and experimentally [14] that a silver film works as a lens to focus near field light sources with subwavelength resolutions. However, such a *super* lens does not function at a frequency other than silver's natural SPP frequency, and since silver only supports TM-polarized SPPs, the source has to be carefully designed to emit *p* waves only [14]. In what follows, we show that our structures can collect both *s* and *p* waves emitted from a source to form an *all-dimensional subwavelength* image [15].

We first performed microwave experiments to demonstrate this idea. We designed a plasmonic metamaterial [picture given in Fig. 3(a)] with unit-cell shown schematically in the right panel of Fig. 3(a). The structure slightly differs from an ideal fractal after optimizations. We then fabricated a series of samples with different thicknesses $H$. In our imaging experiments, we put a dipole antenna on the source plane 1mm above the lenses, and place a receiver dipole antenna on an image plane 1mm below the lenses to measure the field distributions of the images [16]. Both antennas are polarized along the *y* direction, connected to a vector network analyzer (Agilent 8722ES), and working at $f = 2.52$GHz coinciding with the plasmon frequency of the designed lenses for *y* polarization [15]. When the source antenna is put at the unit-cell center, the images formed by two lenses with thicknesses $H = 31.5$mm, $H = 63$mm are depicted in Fig. 3(b) and (c) as open circles, which are in excellent agreements with the corresponding FDTD simulations (solid lines) [17]. Both measurements and FDTD results show that the images focused by our lenses are only ~8mm wide, which are $\sim \lambda/15$ recalling $\lambda \approx 119$mm. In contrast, the images formed without lenses (solid squares) do *not* show any subwavelength resolutions at all. In addition, the field strength is enhanced when a lens is added. The subwavelength resolution and enhanced field strength are two important characteristics of the SPP [1]. Since our lens is not laterally homogeneous, we also performed experiments and FDTD simulations to test the position dependence of the imaging quality, and found that the overall resolution is bounded by the periodicity *a*, which is $\sim \lambda/7$ in the present case [12].

Similar effects can be realized at infrared frequencies using the fractal structure designed for Fig. 2. As a comparison, we also adopted the rectangle-hole structure (same as that for Fig. 2) as a lens to focus light sources. We considered two types of source, i.e., two *x*-polarized dipoles working at 41 THz separated by $1\mu m$ either in *x* direction (case 1) or in *y* direction (case 2). For these two cases, we show the FDTD calculated images formed without any lenses in Fig. 4(a) and 4(d), those with a $0.5\mu m$-thick rectangle-hole structure lens in Fig. 4(b) and 4(e), and those with our fractal structure lens in Fig. 4(c) and 4(f), correspondingly. Here, the source (image) plane is $0.1\mu m$ above (below) the lens [see the right panel in Fig. 1]. Since the sources are located within a subwavelength region, two sources cannot be clearly distinguished without the lens [Fig. 4(a) and 4(d)]. With the rectangle-hole structure lens, case 1 can be distinguished [Fig. 4(b)] but the resolution along *y* direction is not subwavelength. More seriously, this lens cannot distinguish case 2 at all [Fig. 4(e)]. These are all caused by the fact that this structure does not support TE-polarized SPPs (see Fig. 2). With our lens, however, two sources are clearly distinguishable in both cases, and the formed images are *subwavelength along all directions*, with much enhanced field strength (see the E-field scales in Fig. 4).

To explore the underlying physics, we assume the source taking a simple form $\vec{J}(\vec{r},t) = \hat{x} P_0 \delta(\vec{r}) e^{-i\omega t}$, and calculate the EM fields by a Green's function method [18]. We find the fields on the image plane as

$$E_x(x,y) = -\frac{i\mu_0 P_0}{8\pi^2} \int \frac{e^{ik_\parallel(x\cos\phi + y\sin\phi)} e^{-ik_z d}}{k_z} \left[ T^{TE}(k_\parallel) \sin^2\phi + \frac{k_z^2}{k_0^2} T^{TM}(k_\parallel) \cos^2\phi \right] k_\parallel dk_\parallel d\phi \qquad (1)$$

where $k_z = \sqrt{k_0^2 - k_\parallel^2}$ with $k_0 = \omega/c$, *d* is the source-image distance, and $T^{TE}(k_\parallel)$ and $T^{TM}(k_\parallel)$ are the transmission coefficients for TE and TM waves with $\vec{k} = k_z \hat{z} + \vec{k}_\parallel$. To get an *all-dimensional subwavelength* resolution, we need to collect the evanescent components (with $k_\parallel > k_0$) for both TE and TM polarizations. Fortunately, our lens possesses flat SPP bands right at this frequency [see Fig. 2(a)-(b)] for both TE and TM polarizations, so that those evanescent components with $k_0 < k_\parallel < G$ can be transmitted across the lens, aided by such SPPs. The maximum *k* vector of the SPP band *G* sets a natural resolution limit, so that the highest resolution obtained by our plasmonic lens is *a*, which is much less than the working wavelength.

Our mechanism is different from many others [7-8]. In the mechanisms described in Refs. [7] and [8], the operation frequencies depend the lens thickness. In contrast, our working frequency is *independent* of the lens thickness, demonstrated both experimentally and theoretically in Fig. 3. Such a unique property makes our structure a good candidate for far field imaging.

In short, we showed that a metal plate with fractal-shaped holes can be homogenitized as a plamonic metamaterial to support SPPs in both TE and TM polarizations, without using high-index insertions. We demonstrated that it can focus light sources with *all dimensional subwavelength* resolutions at any desired frequency with careful designs. We thank C. T. Chan, M. Qiu and N. Fang for many stimulating discussions. This work was supported by the China-973 Program (2004CB719800), the NSFC (60725417, 60531020, 60671003, 60701007), NCET-07-0750 and PCSIRT.

**References (with paper titles):**

* Corresponding author. E-mail: phzhou@fudan.edu.cn

[1] W. L. Barnes, A. Dereux, T. W. Ebbesen, "Surface plasmon subwavelength optics," Nature **424**, 824 (2003).

[2] T. W. Ebbesen *et al*., "Extraordinary optical transmission through sub-wavelength hole arrays," Nature **391**, 667 (1998).

[3] H. J. Lezec *et. al.* "Beaming light from a subwavelength aperture," Science **297**, 820 (2002).

[4] F. J. Garcia-Vidal *et al.*, "Surfaces with holes in them: new plasmonic metamaterials," J. Opt. A: Pure Appl. Opt. **7**, S97 (2005).

[5] J. B. Pendry, L. Martin-Moreno, F. J. Garcia-Vidal, "Mimicking surface plasmons with structured surfaces," Science **305**, 847 (2004).

[6] Y. Shin, *et al.*, "Evanescent tunneling of an effective surface plasmon excited by convection electrons," Phys. Rev. Lett. **99**, 147402 (2007).

[7] A. Ono, J. Kato, S. Kawata, "Subwavelength optical imaging through a metallic nanorod array," Phys. Rev. Lett. **95**, 267407 (2005).

[8] P. A. Belov, Y. Hao, S. Sudhakaran, "Subwavelength microwave imaging using an array of parallel conducting wires as a lens," Phys. Rev. B **73**, 033108 (2006).

[9] A. Farjadpour *et. al.*, "Improving accuracy by subpixel smoothing in the finite-difference time domain," Opt. Lett. **31**, 2972 (2006).

[10] I. El-Kady *et. al*., "Metallic photonic crystals at optical wavelengths," Phys. Rev. B **62**, 15299 (2000).

[11] H. Raether, Surface Plasmons (ed. G. Hohler), Springer, Berlin, 1988.

[12] S. Xiao *et al.*, unpublished.

[13] J. B. Pendry, "Negative refraction makes a perfect lens," Phys. Rev. Lett. **85**, 3966 (2000).

[14] N. Fang, H. Lee, C. Sun, and X. Zhang, "Sub-diffraction-limited optical imaging with a silver superlens," Science **308**, 534 (2005).

[15] We note that the imaging functionality of the present lens depends on the electric polarization of the source.

[16] In our experiments, we only measured the one-dimensional field distributions along the line perpendicular to the antenna on the image planes.

[17] CONCERTO 7.0, Vector Fields Limited, England, (2008).

[18] L. Zhou and C. T. Chan, "Relaxation mechanisms in three-dimensional

metamaterial lens focusing," Opt. Lett. **30**, 1812 (2005).

**References (without paper titles):**

* Corresponding author. E-mail: phzhou@fudan.edu.cn

[1] W. L. Barnes, A. Dereux, T. W. Ebbesen, Nature **424**, 824 (2003).

[2] T. W. Ebbesen *et. al*., Nature **391**, 667 (1998).

[3] H. J. Lezec *et. al*. Science **297**, 820 (2002).

[4] F. J. Garcia-Vidal *et al.*, J. Opt. A: Pure Appl. Opt. **7**, S97 (2005).

[5] J. B. Pendry, L. Martin-Moreno, F. J. Garcia-Vidal, Science **305**, 847 (2004).

[6] Y. Shin, *et al.*, Phys. Rev. Lett. **99**, 147402 (2007).

[7] A. Ono, J. Kato, S. Kawata, Phys. Rev. Lett. **95**, 267407 (2005).

[8] P. A. Belov, Y. Hao, S. Sudhakaran, Phys. Rev. B **73**, 033108 (2006).

[9] A. Farjadpour *et. al.*, Opt. Lett. **31**, 2972 (2006).

[10] I. El-Kady *et. al*., Phys. Rev. B **62**, 15299 (2000).

[11] H. Raether, Surface Plasmons (ed. G. Hohler), Springer, Berlin, 1988.

[12] S. Xiao *et al.*, unpublished.

[13] J. B. Pendry, Phys. Rev. Lett. **85**, 3966 (2000).

[14] N. Fang, H. Lee, C. Sun, and X. Zhang, Science **308**, 534 (2005).

[15] We note that the imaging functionality of the present lens depends on the electric polarization of the source.

[16] In our experiments, we only measured the one-dimensional field distributions along the line perpendicular to the antenna on the image planes.

[17] CONCERTO 7.0, Vector Fields Limited, England, (2008).

[18] L. Zhou and C. T. Chan, Opt. Lett. **30**, 1812 (2005).

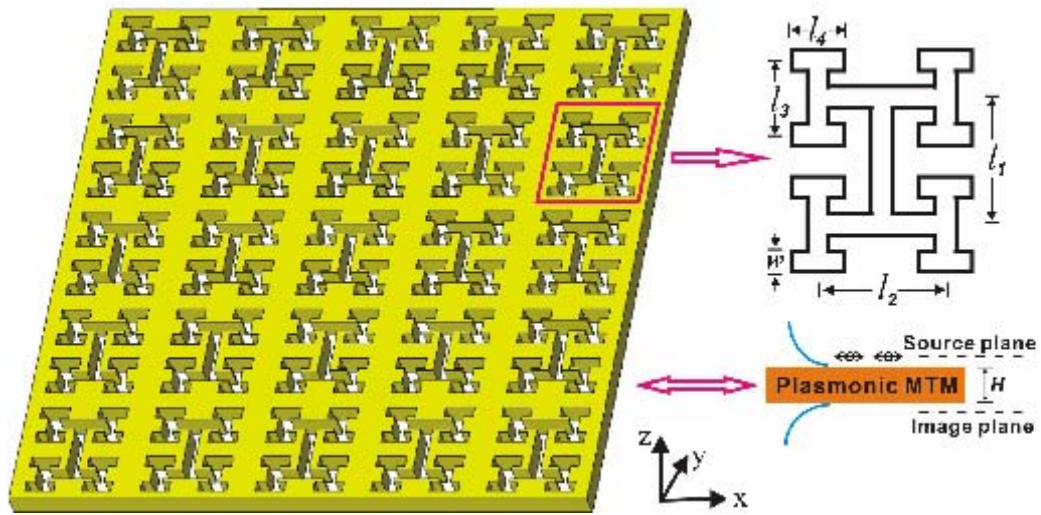

Fig. 1 (Color online) Geometry of the fractal plasmonic metamaterial (MTM). Unit cell strcuture: $l_1 = l_2 = 0.5\mu m,\ l_3 = l_4 = 0.25\mu m,\ w = 0.06\mu m,\ H = 0.5\mu m$.

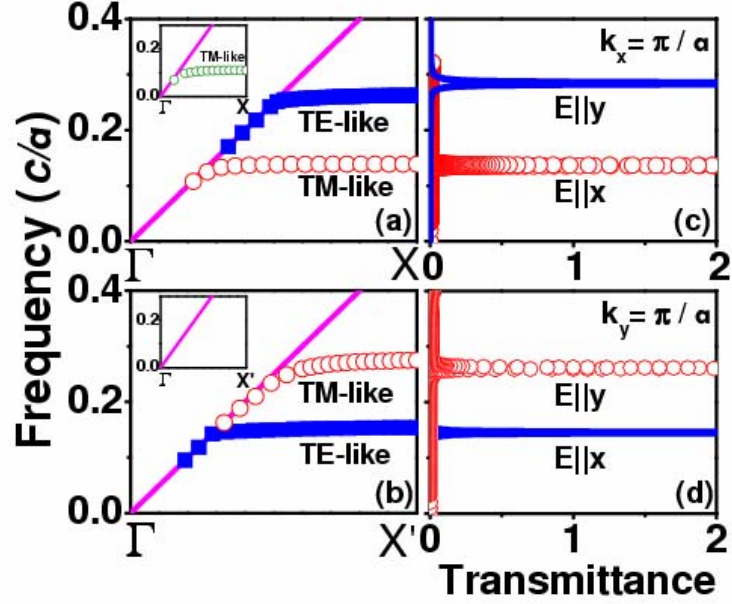

Fig. 2 (Color online) SPP band structrues of the fractal plate calculated by FDTD simulations for (a) $\Gamma \to X$ and (b) $\Gamma \to X'$ directions; Under the conditions of (c) $k_x = \pi/a$ and (d) $k_y = \pi/a$, FDTD calculated transmission spectra under incident plane evanescent waves with different polarizations. Insets: SPP band structrues of rectangle hole plate calculated by FDTD simulations, with structural details $d = 1\mu m, s = 4.2d, a = 0.3d, L = 0.5d$ (see Ref. [6]).

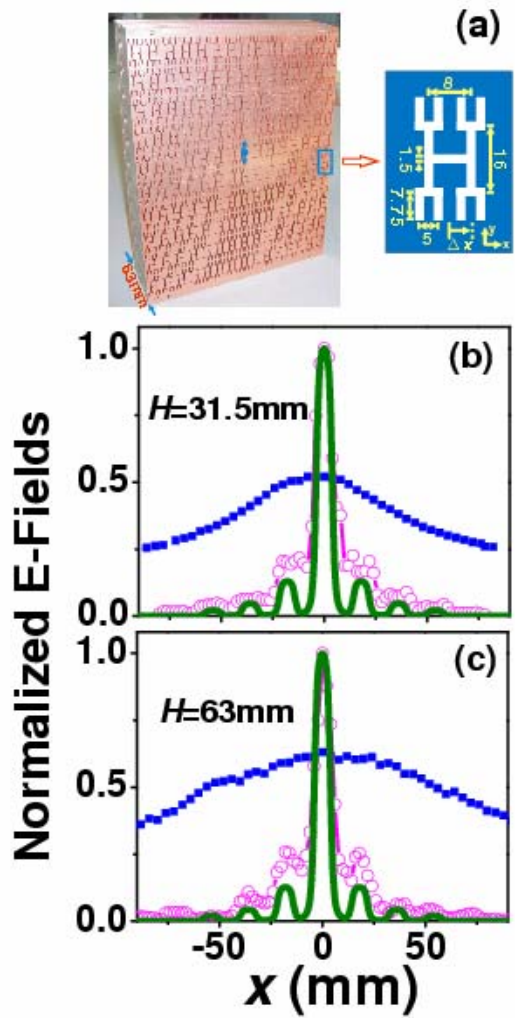

Fig. 3 (Color online) (a) Picture of a 63mm-thick fractal plasmonic metamaterial and its unit cell structure (all lengths are measured in mm). Here the periodicity is 18mm (32mm) in $x$ ($y$) direction. (b)-(c) Electric field distributions along the line perpendicular to the antenna on the image plane obtained by experiments (open circles) and FDTD simulations (solid lines) for different lens thickness, referenced by the experimental results measured without any lens (solid squares). Here, the maximum electric field is normalized to 1 in the presence of a lens.

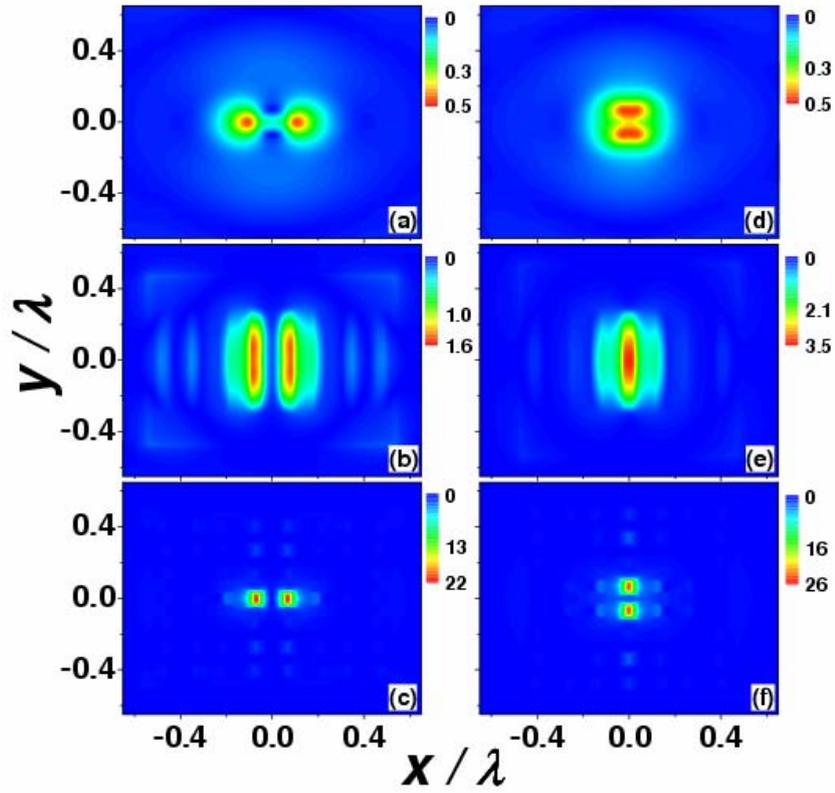

Fig. 4 (Color online) FDTD calculated E-field patterns on the image planes obtained without a lens (a)+(d), with a rectangle-hole structure lens (b)+(e), and with our fractal structure lens (c)+(f). Here, the two *x*-polarized dipole sources are separated by $1\mu m$ in *x* direction for (a), (b), and (c), and in *y* direction for (d), (e) and (f). Both the rectangle-hole structure and fractal structure are the same as those in Fig. 2